\documentclass[sigconf]{acmart}

\setcopyright{acmlicensed}
\copyrightyear{2018}
\acmYear{2018}
\acmDOI{XXXXXXX.XXXXXXX}
\acmConference[Conference acronym 'XX]{Make sure to enter the correct
  conference title from your rights confirmation email}{June 03--05,
  2018}{Woodstock, NY}
\acmISBN{978-1-4503-XXXX-X/2018/06}

\usepackage{graphicx}
\usepackage{booktabs}
\usepackage{subcaption}
\usepackage{multirow}
\usepackage[ruled,vlined]{algorithm2e}
\usepackage{float}
\usepackage{amsmath}
\usepackage{threeparttable}
\usepackage{tabularx}
\usepackage{xspace}
\usepackage[inline]{enumitem}
\usepackage{framed}
\usepackage{booktabs}

\theoremstyle{definition}

\begin{document}

\title{Can LLMs Help Decentralized Dispute Arbitration? \\A Case Study of UMA-Resolved Markets on Polymarket}

\author{Junhao Wen}
\affiliation{%
  \institution{Southwest University}
  \city{Chongqing}
  \country{China}
}
\email{wen264132454@email.swu.edu.cn}

\author{Juncen Zhou}
\affiliation{%
  \institution{University of Sydney}
  \city{Sydney}
  \country{Australia}
}
\email{zhou_2002510@163.com}

\author{Junjie Huang}
\affiliation{%
  \institution{Southwest University}
  \city{Chongqing}
  \country{China}
}
\email{junjiehuang@swu.edu.cn}

\renewcommand{\shortauthors}{Junhao Wen, Juncen Zhou, and Junjie Huang}
\authornote{Junjie Huang is the Corresponding author.}

\begin{abstract}
Web3 prediction markets, exemplified by Polymarket, have gained prominence for leveraging collective intelligence to forecast a wide range of social, political, and sports events. However, among the thousands of prediction market events, consensus \texttt{disputes} still arise due to imperfections in market mechanisms. On Polymarket alone, the trading volume involving disputed events has reached \$972,370,804.71, underscoring the critical need for objective and efficient dispute resolution. In this study, we introduce large language models (LLMs) to: (1) evaluate whether web-enabled LLMs can reproduce the decision quality of UMA’s on-chain voting process once a dispute has been raised, and (2) predict, based on event rules, which market events are likely to face future \texttt{disputes} before they occur. Our findings show that LLMs are unable to reliably predict which events will become disputed in advance; however, once a dispute is initiated, web-enabled LLMs achieve 89.58\% agreement with UMA’s final resolutions and demonstrate strong stability.

\end{abstract}



\maketitle

\section{Introduction}
Decentralized prediction markets have emerged as one of the most prominent applications of Web3, offering a mechanism for aggregating collective intelligence to forecast real-world events. Platforms such as Polymarket allow users to trade on outcomes across political, economic, and sports domains, with market prices reflecting collective expectations grounded in publicly available information~\cite{wolfers2004prediction}. 
Despite their scalability and transparency, these systems remain vulnerable to \texttt{disputes} (i.e., disagreements over event outcomes raised by users or resolvers) caused by ambiguous event wording, inconsistent external reporting, or evolving real-world developments. 
To address such disagreements, Polymarket relies on \textbf{UMA’s Optimistic Oracle (OO)}\footnote{https://uma.xyz}, which escalates contested outcomes to a decentralized token-holder vote. 
While effective in many cases, this governance process has faced criticism for potential subjectivity, coordination challenges, and susceptibility to interpretive ambiguity.
Our data indicates that \texttt{disputed} markets on Polymarket—whether current or historical—represent \$972,370,804.71 in trading volume, demonstrating the vital need for an effective resolution mechanism.

In parallel, recent advances~\cite{chu2024beamaggr,xie2024finben,wei2025instructrag} in large language models (LLMs) have demonstrated strong capabilities in retrieving, interpreting, and synthesizing real-world information. 
Their increasing reliability raises an important question: Can LLMs serve as impartial evaluators in decentralized governance systems, either by predicting which markets are likely to become \texttt{disputed} or by resolving \texttt{disputes} once they occur? 
If effective, LLM-based adjudication could offer a scalable and transparent complement to traditional, human-driven oracle mechanisms.

This study offers the first systematic examination of LLMs within the lifecycle of Web3 prediction-market governance. We investigate two core research questions: 
\textbf{RQ1} evaluates \textbf{whether LLMs can reproduce UMA’s final decisions} in \texttt{disputed} Polymarket events using only information available before the UMA vote is finalized, thereby assessing whether LLMs can effectively substitute for the token-holder voting stage. 
\textbf{RQ2} examines \textbf{whether LLMs can predict \texttt{dispute} formation in advance} based solely on the semantic content of event rules. Our results show that although LLMs struggle to anticipate which events will eventually be \texttt{disputed}, they achieve high pre-vote agreement with UMA’s final outcomes once a \texttt{dispute} has been raised—achieving 89.58\% for DeepSeek V3.1 and 89.19\% for Qwen Max, with Qwen Max further demonstrating exceptionally strong internal stability (96.14\%). These findings indicate that while LLMs exhibit limited predictive capability for identifying \texttt{disputes}, they serve as reliable and consistent evaluators during the \texttt{post-dispute} resolution phase.

Overall, this work highlights both the promise and the limitations of LLMs in decentralized governance systems. Rather than replacing existing oracle mechanisms, LLMs may function as consistent, auditable, and efficient assistants in \texttt{dispute} resolution, helping to enhance transparency and reduce the burdens associated with human-driven voting processes.

\begin{figure}[h!]
    \centering
    \includegraphics[width=1\linewidth]{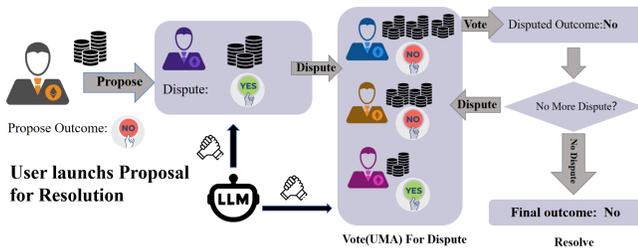}
    \caption{The Polymarket dispute lifecycle: a proposed outcome can be challenged via staking, triggering stake-weighted voting rounds until resolution. We employ LLMs to predict and resolve these \texttt{disputes}.}

    \label{fig:uma}
\end{figure}

\section{Background}
\label{sec:background}
\subsection{Polymarket}
\textbf{Polymarket}\footnote{https://polymarket.com} is a leading decentralized prediction market platform that enables users to trade on the outcomes of real-world events across politics, economics, and culture. It leverages market mechanisms to aggregate public information and quantify collective beliefs through prices that reflect event probabilities. Built on polygon blockchain technology, Polymarket ensures transparency, liquidity, and censorship resistance, attracting significant trading volumes and mainstream attention. To securely resolve market outcomes, it relies on \textbf{UMA’s Optimistic Oracle (OO)} for decentralized and verifiable truth determination.

\subsection{UMA}
UMA (short for \textit{Universal Market Access}) is a decentralized protocol deployed on Ethereum and other blockchains. One of its core components is the \textbf{Optimistic Oracle (OO)}, a mechanism designed to securely and efficiently bring verifiable real-world data or event outcomes (e.g., \textit{whether a person won a competition}). 

The \textbf{OO} operates under an \textsc{assume honesty first} paradigm: when a participant submits an \textit{assertion} about a real-world fact, they must post a bond as collateral, and the claim then enters a predetermined \textit{challenge period}. 
If no one \texttt{disputes} it during this window, the assertion is considered correct by default and finalized automatically. 
However, if challenged, the claim escalates to a \texttt{dispute phase}, where UMA’s \href{https://docs.uma.xyz/protocol-overview/dvm-2.0}\textbf{Data Verification Mechanism (DVM)} based dispute resolution life circle  is invoked(see Figure~\ref{fig:uma}). 
The \textbf{DVM} relies on UMA token holders to vote on the correct outcome, aligning economic incentives to ensure honest and accurate resolution. 
This design allows most queries to be resolved efficiently and inexpensively, while only the small fraction of \texttt{disputed} cases invoke the more resource-intensive voting process, striking a balance between decentralization and scalability.

Nevertheless, the \textbf{OO} system also carries certain risks and limitations. 
Its optimistic design relies on rational and well-incentivized participants; if voting power becomes overly concentrated or if community engagement is low, outcomes could be influenced by a small subset of token holders. 
Moreover, because many event statements are expressed in natural language, ambiguity in question phrasing can lead to inconsistent interpretations or contentious resolutions. 
These risks have manifested in real-world applications such as Polymarket.

\subsection{\texttt{Disputed} Example}
A notable example of these risks occurred in the Polymarket event \href{https://polymarket.com/event/will-zelenskyy-wear-a-suit-before-july}{\textit{Will Zelenskyy wear a suit before July?}}, which drew over \$200 million in trading volume. 
The market asked whether Ukrainian President Volodymyr Zelenskyy would appear publicly in a ``suit'' between May 22 and June 30, 2025. 
Controversy emerged after he attended a NATO summit wearing a dark jacket, shirt, and trousers, which some participants described as a suit while others rejected because it lacked traditional elements such as a matching set or a tie.
Both ``Yes'' and ``No'' outcomes were proposed and \texttt{disputed}, prompting \textbf{OO} to resolve the question through its voting mechanism. 
The final decision of ``No'' triggered backlash  ~\cite{zelensky_suit_polymarket} from traders who claimed governance centralization and subjective interpretation had compromised fairness. 
UMA’s co-founder, Hart Lambur, rejected manipulation claims, but the controversy revealed how even simple, natural-language questions can challenge decentralized truth systems reliant on collective judgment.

These controversies highlight that while \textbf{OO} represents a groundbreaking innovation in decentralized data verification, challenges remain in governance, incentive alignment, and the interpretation of natural-language conditions.

\subsection{Related Work}
Prediction markets have long been studied as mechanisms for aggregating distributed beliefs, with foundational work on automated market makers, illustrating how liquidity and incentives shape information aggregation ~\cite{othman2013practical,
chen2010survey}. 
Decentralized variants, such as Polymarket, extend these ideas to trust-minimized environments, relying on oracle-based governance to finalize outcomes ~\cite{chiarelli2023blockchainoracles,kk2021decentralized}. 
UMA's Optimistic Oracle represents a prominent design in this space, using challenge-based verification and token-holder voting to resolve ambiguous or \texttt{disputed} events ~\cite{uma2024optimistic}. Despite their scalability, existing oracle systems also face critical limitations. For example, implementation surveys have shown that trust-minimized oracle designs cannot fully eliminate integrity and freshness risks, especially when aggregating data from heterogeneous sources ~\cite{pasdar2023survey}.

Concurrently, rapid advances in large language models (LLMs) have spurred research into autonomous agents and LLMs' web-search for decentralized governance~\cite{liu2025dellma,peter2025decentralising,spatharioti2025effects}. 
Recent work shows that web-enabled LLM agents can evaluate DAO proposals and align with collective voting behavior~\cite{daoai2025}, while self-sovereign LLM agents demonstrate the feasibility of trust-minimized autonomous decision-makers in Web3 systems ~\cite{deagent2025}. 
Complementary studies in AI-assisted governance and \texttt{dispute} resolution~\cite{savelka2023large,bommarito2024lawgpt} further suggest that LLMs can perform structured fact interpretation and rule-based reasoning. 
These findings collectively indicate that LLM-driven agents may augment or partially automate governance workflows in decentralized environments.

In this paper, we present the first attempt to apply large language models to decentralized dispute arbitration and investigate whether LLMs can assist in resolving on-chain disputes.

\begin{figure}[htbp]
\centering
\includegraphics[width=1.0\linewidth]{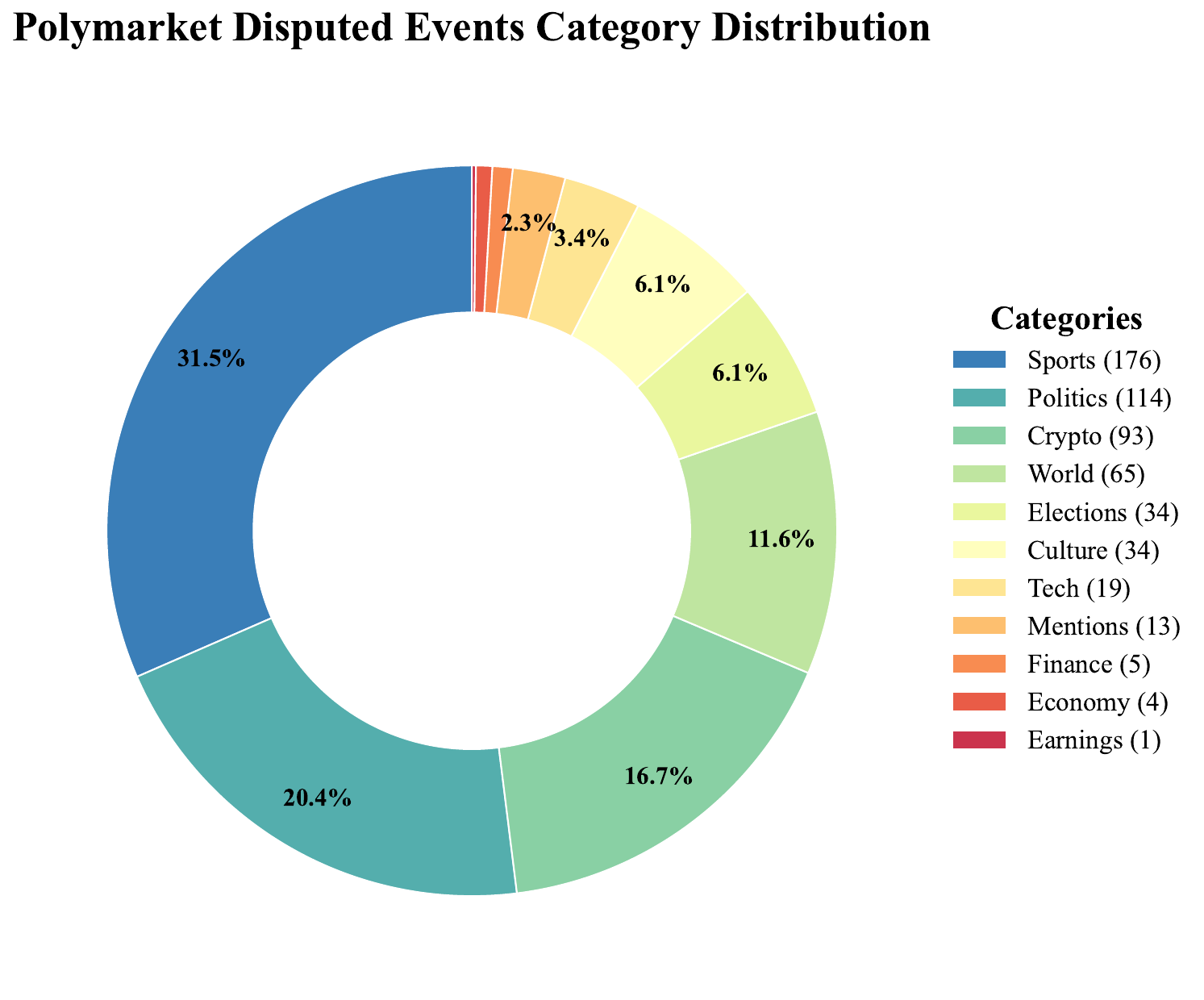}
\caption{Distribution of Polymarket disputed Events by Category. \textbf{Sports} (31.5\%), \textbf{Politics} (20.4\%), and \textbf{Crypto} (16.7\%) are the three largest categories.}
\label{fig:disputed-category}
\end{figure}

\section{Dataset and Data Preprocessing}
We retrieved all available event data from Polymarket using the official \textbf{API}\footnote{https://gamma-api.polymarket.com/markets} endpoint, iteratively accessing paginated results to ensure complete coverage of all active and historical markets.
The API was queried three times at regular intervals, and we observed that each subsequent query returned all previously obtained events along with newly created ones, indicating that the official endpoint is  stable and consistent\footnote{The most recent retrieval was performed on UTC \texttt{2025-10-27 11:58:47}, 
yielding a total of 140,582 Polymarket events. }. Each record represents a prediction market event with detailed metadata including its identifier, 
question content, \texttt{resolution statuses}, market outcomes, prices, and sub-event grouping information.Using the \texttt{umaResolutionStatuses} , we filtered for records containing the keyword \texttt{disputed}, resulting in a total of 558 user-disputed events.Based on these events, we will conduct further filtering and analysis in Section~\ref{sec:nextsteps} and Section~\ref{sec:nextsteps2}.

A multi-LLM classification employing three distinct models was performed on the disputed events. The outcome demonstrated high classification stability, with $457$ events ($81.90\%$) achieving full consistency (\textit{Very Stable}) and $98$ events ($17.56\%$) showing large consistency (\textit{Stable}), resulting in nearly $99.5\%$ of events exhibiting robust agreement. Only three events ($0.54\%$) were inconsistent (\textit{Unstable}), which were subsequently resolved through manual review and assigned to the \textbf{Mentions} category (IDs: $542935$, $534923$, $570937$). As depicted in Figure~\ref{fig:disputed-category}, the distribution of event categories post-classification reveals a significant concentration of disputed events within the \textbf{Sports}, \textbf{Politics}, and \textbf{Crypto} domains, collectively accounting for approximately $68.6\%$ of all disputes. This pronounced asymmetry, reliably identified via highly stable LLM classification, suggests a clear direction for subsequent research, primarily focusing on characterizing the unique factors driving disputes specifically within these three predominant categories.

\section{Can LLMs Reproduce UMA’s Final Decisions?}\label{sec:nextsteps}

A Polymarket event’s complete \texttt{dispute} trajectory (see Figure~\ref{fig:uma}) is encoded in the \texttt{umaResolutionStatuses} field, which records each state transition the market undergoes. A typical sequence may look like:
\[
\texttt{["proposed", "disputed", "proposed", "resolved"]}.
\]
Using this field in our collected API data, we first identify all individual occurrences of the \texttt{disputed} state, yielding a total of 558 user-initiated dispute events. Next, we restrict attention to markets that both (1) contain at least one \texttt{disputed} transition and (2) ultimately reach a terminal \texttt{resolved} state. 
Applying these criteria results in a final set of 259 \texttt{disputed-and-resolved} markets, which form the basis of our subsequent analysis.Our goal is to investigate whether LLMs equipped with web-search capabilities can independently reproduce UMA's final resolutions. 
If LLMs can match UMA' outcomes, this would suggest the feasibility of a more transparent and objective mechanism for resolving decentralized prediction markets—potentially mitigating subjective biases or 
manipulation risks in the UMA voting process.To evaluate this, we prompt five frontier LLMs to 
search only for information available before the UMA decision timestamp and to infer 
the correct market resolution according to the official Polymarket rulebook. 
Each model is  queried three independent times to measure stability. 
For each event, a model’s final answer  is determined by majority vote across its three runs. 
Consistency is then computed as the agreement rate between the model’s majority-vote prediction and the ground-truth UMA resolution across the 259 \texttt{disputed} markets.

The results, including both model--UMA consistency and model internal stability 
(very-stable, stable, unstable across the three runs), are shown in Table~\ref{tab:llm_stability}.
Across the five evaluated frontier models, we observe a clear separation in both alignment with UMA outcomes and internal stability. DeepSeekV3.1, Qwen Max, and Claude-4.5-Sonnet achieve the highest consistency (88--90\%) and very-stable rates (exceeding 95\% for the latter two), demonstrating that frontier LLMs can reproduce UMA’s final resolutions with high fidelity when limited to information available at decision time. While the two OpenAI variants show comparatively lower consistency---especially \texttt{gpt-4o-search-preview} (71.43\%)---their performance still exceeds chance by a wide margin, indicating nontrivial signal extraction under temporal constraints. Overall, these results affirm that LLMs are capable of approximating UMA arbitration outcomes, though persistent model-specific reliability gaps highlight the need for careful validation before deploying LLM-assisted adjudication in decentralized governance settings.

\begin{table}
\centering
\caption{
Consistency reflects how closely LLM predictions align with UMA’s final resolutions across 259 events.
}
\label{tab:llm_stability}
\resizebox{1.0\linewidth}{!}{
\begin{tabular}{lcccc}
\toprule
\textbf{Model} & \textbf{Consistency (259)} & \textbf{Very-Stable} & \textbf{Stable} & \textbf{Unstable} \\
\midrule
DeepSeekV3.1 & 232 (89.58\%) & 227 (87.64\%) & 32 (12.36\%) & 0 (0.00\%) \\

Qwen Max & 231 (89.19\%) & 249 (96.14\%) & 10 (3.86\%) & 0 (0.00\%) \\

Claude-4.5-Sonnet & 228 (88.03\%) & 248 (95.75\%) & 11 (4.25\%) & 0 (0.00\%) \\

gpt4o & 203 (78.38\%) & 166 (64.09\%) & 89 (34.36\%) & 4 (1.54\%) \\

gpt-4o-search-preview & 185 (71.43\%) & 223 (86.10\%) & 35 (13.51\%) & 1 (0.39\%) \\
\bottomrule
\end{tabular}
}

\end{table}

\clearpage
\section{Can LLMs Predict UMA Disputes in Advance?}\label{sec:nextsteps2}

\begin{table}
\centering
\caption{Model performance when predicting dispute outcomes using only pre-dispute semantic information.}
\label{tab:model-results}
\resizebox{\linewidth}{!}{
\begin{tabular}{lcccc}
\toprule
\textbf{Model} & \textbf{Accuracy} & \textbf{Precision} & \textbf{Recall} & \textbf{F1 Score} \\
\midrule
Claude-Sonnet-4.5 & 0.5701 & 0.6305 & 0.3388 & 0.4408 \\
DeepSeek-V3.1 & 0.5410 & 0.5680 & 0.3424 & 0.4273 \\
Qwen-Max & 0.4731 & 0.4601 & 0.3157 & 0.3745 \\
GPT-4o & 0.3998 & 0.3618 & 0.2623 & 0.3041 \\
\bottomrule
\end{tabular}
}
\end{table}

To evaluate this hypothesis, we construct a structured few-shot prompt using twelve events: nine \texttt{disputed} cases covering diverse \texttt{dispute} types (semantic ambiguity, definitional uncertainty, evidentiary limitations, reporting inconsistencies, and temporal instability), plus three non-disputed events. 
The choice of twelve exemplars balances coverage of major dispute categories while keeping the prompt size tractable for all models. 
Models receive only the event rule text and timestamps, ensuring that predictions rely solely on semantic features of the rule descriptions.

The dataset for this experiment is derived from all Polymarket events that contain at least one user-initiated \texttt{disputed} transition, yielding 558 total dispute events. 
From these, nine \texttt{disputed} events are selected as in-context exemplars for the prompt, leaving 549 remaining \texttt{disputed} cases. 
For evaluation, we pair these 549 cases with 549 randomly sampled non-disputed events, forming a balanced test set of 1,098 samples. 
Because the dataset is balanced, a random or majority-class baseline corresponds to an accuracy of 0.50.
We evaluate four LLMs on this test set, and their performance is shown in Table~\ref{tab:model-results}.

From Table~\ref{tab:model-results}, we observe that predictive performance remains limited across all models. 
Although some models perform slightly above the random baseline (for example, Claude with an accuracy of 0.5701), none achieve strong discriminative ability. This indicates that semantic information derived solely from event rules and pre-dispute descriptions provides only weak signals for forecasting future \texttt{disputes}.

This observation leads to two insights.
First, within our dataset, neither the clarity nor the structural quality of Polymarket rule wording appears to play a decisive role in determining whether an event will later become \texttt{disputed}. Textual cues alone are not sufficient for reliably anticipating \texttt{dispute} formation.
Second, the emergence of \texttt{disputes} seems to be shaped mainly by external real-world developments, including evolving facts, unstable or incomplete evidence, conflicting reports, media dynamics, and delays in authoritative confirmation.

\section{Conclusion}

Our analysis provides clear answers to the two research questions posed in this work.
For \textbf{RQ1}, we find that advanced LLMs with controlled access to public information can closely align with UMA’s final dispute resolution outcomes, showing strong agreement and internal stability within our experimental setting.
This result indicates that the evidence interpretation process, which is traditionally carried out through decentralized voting by token holders, can in part be approximated by an autonomous agent equipped with retrieval capabilities. Rather than replacing human governance, LLMs may function as complementary adjudication tools that deliver consistent and auditable reasoning and help promote transparency and scalability in optimistic oracle frameworks.

For \textbf{RQ2}, we show that LLMs cannot reliably predict which Polymarket events will become \texttt{disputed} when limited to the semantic content of event rules. Textual clarity or ambiguity in the rule descriptions does not provide a sufficient signal for forecasting \texttt{dispute} formation. 
Instead, disputes tend to emerge from external real-world uncertainty, evolving evidence, reporting inconsistencies, and delays in authoritative confirmation, all of which fall outside the rule text itself.

Taken together, these findings point to a natural division of labor. The emergence of \texttt{disputes} is shaped by the evolving state of real-world information, whereas \texttt{dispute} resolution depends on interpreting that information once it becomes available. Within this lifecycle, LLMs appear well-suited for assisting in the resolution stage but are not effective for anticipating \texttt{disputes} based solely on event semantics. This reveals a realistic and bounded role for LLM-based agents in decentralized governance: providing stable and consistent evaluations in the post-dispute stage while leaving pre-dispute forecasting to market behavior.

\bibliographystyle{ACM-Reference-Format}
\bibliography{refs}


\begin{thebibliography}{18}


\ifx \showCODEN    \undefined \def \showCODEN     #1{\unskip}     \fi
\ifx \showISBNx    \undefined \def \showISBNx     #1{\unskip}     \fi
\ifx \showISBNxiii \undefined \def \showISBNxiii  #1{\unskip}     \fi
\ifx \showISSN     \undefined \def \showISSN      #1{\unskip}     \fi
\ifx \showLCCN     \undefined \def \showLCCN      #1{\unskip}     \fi
\ifx \shownote     \undefined \def \shownote      #1{#1}          \fi
\ifx \showarticletitle \undefined \def \showarticletitle #1{#1}   \fi
\ifx \showURL      \undefined \def \showURL       {\relax}        \fi
\providecommand\bibfield[2]{#2}
\providecommand\bibinfo[2]{#2}
\providecommand\natexlab[1]{#1}
\providecommand\showeprint[2][]{arXiv:#2}

\bibitem[Capponi et~al\mbox{.}(2025)]%
        {daoai2025}
\bibfield{author}{\bibinfo{person}{Agostino Capponi} {et~al\mbox{.}}}
  \bibinfo{year}{2025}\natexlab{}.
\newblock \showarticletitle{{DAO-AI}: Evaluating Collective Decision-Making
  with Web-Enabled {LLM} Agents}.
\newblock \bibinfo{journal}{\emph{arXiv preprint arXiv:2510.21117}}
  (\bibinfo{year}{2025}).
\newblock


\bibitem[Chen and Pennock(2010)]%
        {chen2010survey}
\bibfield{author}{\bibinfo{person}{Yiling Chen} {and} \bibinfo{person}{David
  Pennock}.} \bibinfo{year}{2010}\natexlab{}.
\newblock \showarticletitle{A survey of prediction market design}. In
  \bibinfo{booktitle}{\emph{Algorithmic Game Theory}}. \bibinfo{pages}{1--33}.
\newblock


\bibitem[Chiarelli et~al\mbox{.}(2023)]%
        {chiarelli2023blockchainoracles}
\bibfield{author}{\bibinfo{person}{Andrea Chiarelli} {et~al\mbox{.}}}
  \bibinfo{year}{2023}\natexlab{}.
\newblock \showarticletitle{A Systematic Literature Review of Blockchain
  Oracles}.
\newblock \bibinfo{journal}{\emph{Aalto University Publication Series}}
  (\bibinfo{year}{2023}).
\newblock


\bibitem[Chu et~al\mbox{.}(2024)]%
        {chu2024beamaggr}
\bibfield{author}{\bibinfo{person}{Zheng Chu}, \bibinfo{person}{Jingchang
  Chen}, \bibinfo{person}{Qianglong Chen}, \bibinfo{person}{Haotian Wang},
  \bibinfo{person}{Kun Zhu}, \bibinfo{person}{Xiyuan Du},
  \bibinfo{person}{Weijiang Yu}, \bibinfo{person}{Ming Liu}, {and}
  \bibinfo{person}{Bing Qin}.} \bibinfo{year}{2024}\natexlab{}.
\newblock \showarticletitle{{B}eam{A}gg{R}: Beam Aggregation Reasoning over
  Multi-source Knowledge for Multi-hop Question Answering}. In
  \bibinfo{booktitle}{\emph{ACL}}. \bibinfo{pages}{1229--1248}.
\newblock


\bibitem[Hu et~al\mbox{.}(2025)]%
        {deagent2025}
\bibfield{author}{\bibinfo{person}{Botao~Amber Hu}, \bibinfo{person}{Yuhan
  Liu}, {and} \bibinfo{person}{Helena Rong}.} \bibinfo{year}{2025}\natexlab{}.
\newblock \showarticletitle{Trustless Autonomy: Self-Sovereign Large Language
  Model Agents in Decentralized Systems}.
\newblock \bibinfo{journal}{\emph{arXiv preprint arXiv:2505.09757}}
  (\bibinfo{year}{2025}).
\newblock


\bibitem[Khalili and Knibbs(2025)]%
        {zelensky_suit_polymarket}
\bibfield{author}{\bibinfo{person}{Joel Khalili} {and} \bibinfo{person}{Kate
  Knibbs}.} \bibinfo{year}{2025}\natexlab{}.
\newblock \bibinfo{booktitle}{\emph{Volodymyr Zelensky’s Clothing Has Sparked
  a {Polymarket} Rebellion}}.
\newblock
\urldef\tempurl%
\url{https://www.wired.com/story/volodymyr-zelensky-suit-polymarket-rebellion}
\showURL{%
\tempurl}


\bibitem[Kumar and Khan(2021)]%
        {kk2021decentralized}
\bibfield{author}{\bibinfo{person}{Kartikay Kumar} {and}
  \bibinfo{person}{Muhammad Khan}.} \bibinfo{year}{2021}\natexlab{}.
\newblock \showarticletitle{Decentralized oracles: A comprehensive survey}.
\newblock \bibinfo{journal}{\emph{IEEE Access}}  \bibinfo{volume}{9}
  (\bibinfo{year}{2021}), \bibinfo{pages}{92272--92294}.
\newblock


\bibitem[Liu et~al\mbox{.}(2025)]%
        {liu2025dellma}
\bibfield{author}{\bibinfo{person}{Ollie Liu}, \bibinfo{person}{Deqing Fu},
  \bibinfo{person}{Dani Yogatama}, {and} \bibinfo{person}{Willie Neiswanger}.}
  \bibinfo{year}{2025}\natexlab{}.
\newblock \showarticletitle{{DeLLMa}: Decision Making Under Uncertainty with
  Large Language Models}. In \bibinfo{booktitle}{\emph{ICLR}}.
\newblock


\bibitem[Othman et~al\mbox{.}(2013)]%
        {othman2013practical}
\bibfield{author}{\bibinfo{person}{Othman} {et~al\mbox{.}}}
  \bibinfo{year}{2013}\natexlab{}.
\newblock \showarticletitle{A practical liquidity-sensitive automated market
  maker}.
\newblock \bibinfo{journal}{\emph{TEAC}} \bibinfo{volume}{1},
  \bibinfo{number}{3} (\bibinfo{year}{2013}), \bibinfo{pages}{1--25}.
\newblock


\bibitem[Pasdar and Lee(2023)]%
        {pasdar2023survey}
\bibfield{author}{\bibinfo{person}{Amir Pasdar} {and}
  \bibinfo{person}{Young~Choon Lee}.} \bibinfo{year}{2023}\natexlab{}.
\newblock \showarticletitle{A Survey on Blockchain Oracle Implementation}.
\newblock \bibinfo{journal}{\emph{Comput. Surveys}} \bibinfo{volume}{55},
  \bibinfo{number}{12} (\bibinfo{year}{2023}), \bibinfo{pages}{1--36}.
\newblock


\bibitem[Peter and Devlin(2025)]%
        {peter2025decentralising}
\bibfield{author}{\bibinfo{person}{Oriane Peter} {and} \bibinfo{person}{Kate
  Devlin}.} \bibinfo{year}{2025}\natexlab{}.
\newblock \showarticletitle{Decentralising {LLM} Alignment: A Case for Context,
  Pluralism, and Participation}. In \bibinfo{booktitle}{\emph{AAAI}}.
\newblock


\bibitem[Savelka et~al\mbox{.}(2023)]%
        {savelka2023large}
\bibfield{author}{\bibinfo{person}{Jaromir Savelka} {et~al\mbox{.}}}
  \bibinfo{year}{2023}\natexlab{}.
\newblock \showarticletitle{Large Language Models in Legal Reasoning: A Study
  on Real-World Case Interpretation}.
\newblock \bibinfo{journal}{\emph{Journal of Artificial Intelligence and Law}}
  (\bibinfo{year}{2023}).
\newblock


\bibitem[Spatharioti et~al\mbox{.}(2025)]%
        {spatharioti2025effects}
\bibfield{author}{\bibinfo{person}{Sofia~Eleni Spatharioti} {et~al\mbox{.}}}
  \bibinfo{year}{2025}\natexlab{}.
\newblock \showarticletitle{Effects of {LLM}-based Search on Decision Making:
  Speed, Accuracy, and Overreliance}. In \bibinfo{booktitle}{\emph{CHI}}.
  \bibinfo{publisher}{ACM}.
\newblock


\bibitem[{UMA Protocol}(2024)]%
        {uma2024optimistic}
\bibfield{author}{\bibinfo{person}{{UMA Protocol}}.}
  \bibinfo{year}{2024}\natexlab{}.
\newblock \bibinfo{title}{Understanding {UMA}'s Optimistic Oracle and Its
  Governance Mechanisms}.
\newblock


\bibitem[Wei et~al\mbox{.}(2025)]%
        {wei2025instructrag}
\bibfield{author}{\bibinfo{person}{Zhepei Wei}, \bibinfo{person}{Wei-Lin Chen},
  {and} \bibinfo{person}{Yu Meng}.} \bibinfo{year}{2025}\natexlab{}.
\newblock \showarticletitle{Instruct{RAG}: Instructing Retrieval-Augmented
  Generation via Self-Synthesized Rationales}. In
  \bibinfo{booktitle}{\emph{ICLR}}.
\newblock


\bibitem[Wolfers and Zitzewitz(2004)]%
        {wolfers2004prediction}
\bibfield{author}{\bibinfo{person}{Justin Wolfers} {and} \bibinfo{person}{Eric
  Zitzewitz}.} \bibinfo{year}{2004}\natexlab{}.
\newblock \showarticletitle{Prediction Markets}.
\newblock \bibinfo{journal}{\emph{Journal of Economic Perspectives}}
  \bibinfo{volume}{18}, \bibinfo{number}{2} (\bibinfo{year}{2004}),
  \bibinfo{pages}{107--126}.
\newblock


\bibitem[Xie et~al\mbox{.}(2024)]%
        {xie2024finben}
\bibfield{author}{\bibinfo{person}{Qianqian Xie} {et~al\mbox{.}}}
  \bibinfo{year}{2024}\natexlab{}.
\newblock \showarticletitle{Fin{B}en: A Holistic Financial Benchmark for Large
  Language Models}. In \bibinfo{booktitle}{\emph{NeurIPS}}.
\newblock


\bibitem[Yao et~al\mbox{.}(2024)]%
        {bommarito2024lawgpt}
\bibfield{author}{\bibinfo{person}{Shunyu Yao} {et~al\mbox{.}}}
  \bibinfo{year}{2024}\natexlab{}.
\newblock \showarticletitle{{Lawyer GPT}: A Legal Large Language Model with
  Enhanced Domain Knowledge and Reasoning Capabilities}. In
  \bibinfo{booktitle}{\emph{Proceedings of the 2024 3rd International Symposium
  on Robotics, Artificial Intelligence and Information Engineering}}.
  \bibinfo{pages}{108--112}.
\newblock


\end{thebibliography}

\end{document}